\documentclass{aa}
\usepackage{graphicx}
\begin{document}
\title{Numerical Simulation of the Surface Flow on the Companion Star in a Close 
Binary System}

\subtitle{II. Construction of Doppler maps and application to Galactic supersoft 
X-ray sources}

\author{Kazutaka Oka\inst{1}, Takuya Matsuda\inst{1}, Izumi Hachisu\inst{2}
\and Henri M.J. Boffin\inst{3} }

\institute{Department of Earth and Planetary Sciences, Kobe University, Kobe,  
657-8501, Japan\\
    \email{kazutaka@kobe-u.ac.jp, tmatsuda@kobe-u.ac.jp}
    \and
    Department of Earth Science and Astronomy, College of Arts and 
    Sciences, University of Tokyo, Komaba, Meguro-ku, Tokyo 153-8902, Japan\\
    \email{hachisu@chianti.c.u-tokyo.ac.jp}
    \and
    European Southern Observatory, Karl-Schwarzschild-Str. 2, D-85748 
Garching-bei-M\"unchen, Germany\\
    \email{hboffin@eso.org}
  }


\abstract{
We perform three-dimensional numerical simulations of surface flows on the 
companion star in a semi-detached binary system and construct the corresponding 
Doppler maps. The mass ratio of the binary system, $q=M_2/M_1$, considered here 
is $q=0.05, 0.33, 0.5, 1, 2$, and $3$. For all cases, we obtain the H-, L1-, and 
L2-eddies, as found in our previous work, and confirm that the 
flow pattern does not heavily depend on the mass ratio.  We propose that this kind of problem be dubbed ``{\it stellar meteorology}.''
The Doppler maps at the position of the companion show a structure tilted towards
clockwise direction and presenting deviations from the critical Roche surface due to the 
L1-eddy and the L2-eddy on the companion star.

    We apply our results to the Galactic supersoft X-ray source RX J0019.8+2156 and try to attribute the low radial velocity component of the 
emission lines of He II ${\lambda}4686$ observed recently to the irradiated spot 
on the surface of the companion rather than that of the white dwarf or the 
accretion disc. Based on the comparison between the observations and our 
constructed Doppler map, we estimate the mass of the companion star in RX 
J0019.8+2156 to be $\sim 2 M_{\odot}$ assuming the mass of the white dwarf star to be 
around $0.6 M_{\odot}$.

   }

\titlerunning{Numerical Simulation of the Surface Flow on a Companion Star
II}  \authorrunning{Oka et al.} 
\maketitle
%

\section{Introduction}

A semi-detached binary system consists of a Roche-lobe-filling companion
star and a (compact) accreting object. Part of the gas of the companion star
is gravitationally attracted by the primary, flows through the L1 point
towards the primary star and, if the latter is smaller than the circularisation radius, forms an accretion disc around it.

Many studies on the accretion disc itself exist, but until recently,
only a very few studies considered the surface flow of the companion,
except for the
pioneering work by Lubow \& Shu (\cite{lub75}). A possible reason for the lack of
studies on the companion is that there had been no effective way to observe
the surface of the companion star as well as the flow on it.

However, Davey \& Smith (\cite{dav92}) demonstrated the surface
mapping method of the companion surface in cataclysmic variables (CVs) using Na I line absorption.
Dhillon \& Watson (\cite{dhi00}) demonstrated a technique of imaging
companion stars of CVs using Roche tomography. Besides, Doppler map
technique also can potentially give some trace of the
surface flow (as considered in this paper).

Oka et al. (\cite{oka02}) performed three-dimensional simulations
of surface flows on the companion star and predicted the existence of
three kinds of eddy associated with a high/low pressure on the companion
star: the H-, L1-, and L2-eddies as schematically shown in Fig.1.
The notations H, L1 and L2 denote the high pressure around a (north) pole,
the low pressure around the L1 point and the low pressure at the opposite
side to the L1 point, respectively.

In a rotating fluid, the pressure gradient force balances the Coriolis
force, and the wind blows thus along isobaric lines. On the Earth, such a
flow is called the geostrophic wind, while on a star it was called the astrostrophic
wind by Lubow \& Shu (\cite{lub75}). Therefore,  a clockwise-rotating eddy 
forms around a high pressure region, while a counter-clockwise rotating eddy appears
around a low pressure region on the northern hemisphere in our
counterclockwise rotating system (viewed from the north). We dub this kind
of problem ``\textit{stellar meteorology}.''

With the observational progress mentioned above as well as theoretical ones in hand, we are now
in the state of investigating the flow on the surface of the companion star in
some more detail.

Oka et al. (\cite{oka02}) considered only the mass ratio of unity.
In reality, however, the range of mass ratio is wide, and it is thus
interesting to investigate the dependence of
the flow pattern on the mass ratio, $q=M_2/M_1$. In the present study, we
consider six cases for the mass ratio: $q=0.05, 0.33, 0.5, 1, 2$, and $3$. As
described above, a Doppler map can potentially give some traces of the
surface flow, and so we construct Doppler maps in the present study as
well.

\begin{figure}
\centering  
\resizebox{\hsize}{!}{\includegraphics{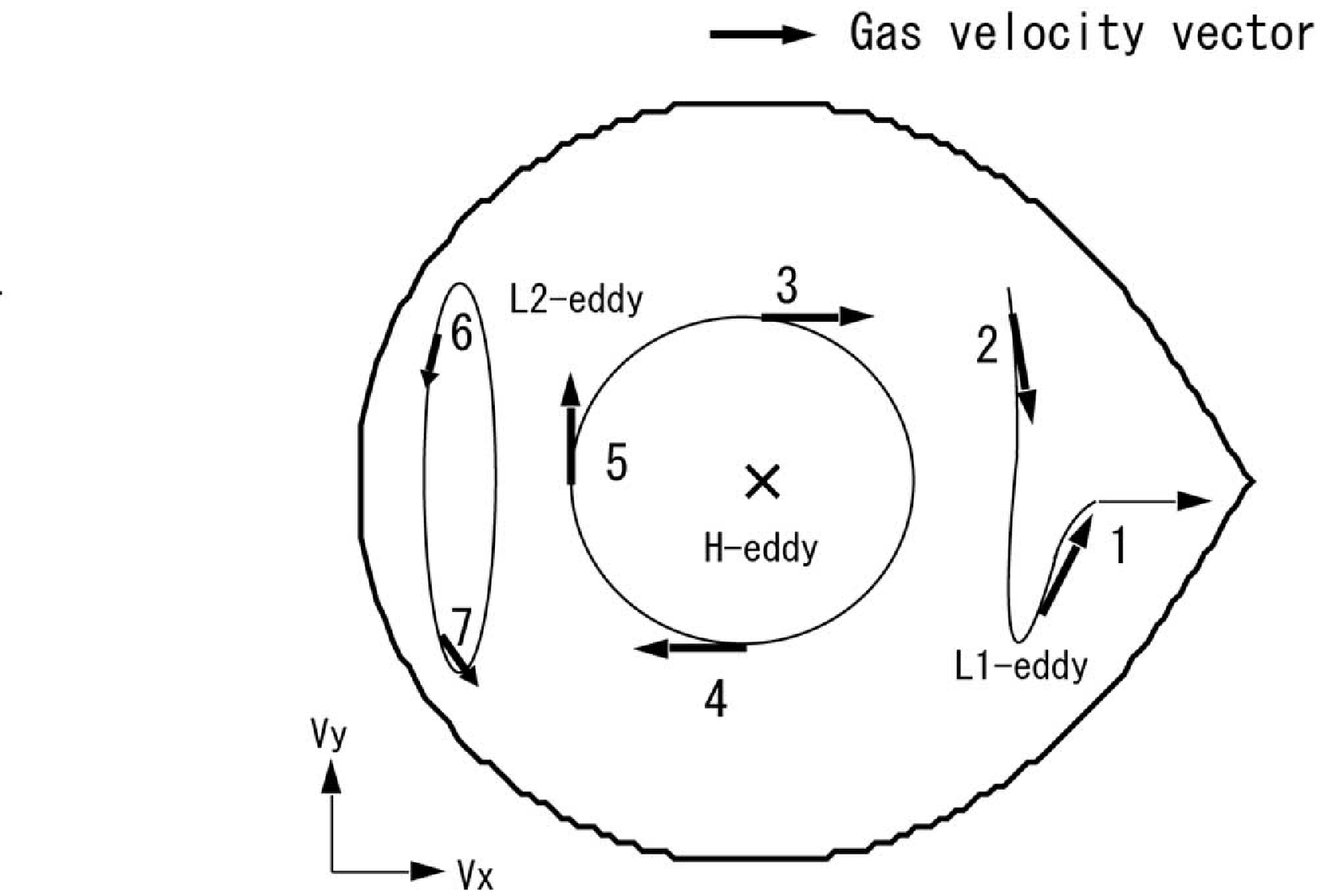}}  
\caption{A schematic diagram of the surface flow of a companion star in a
semi-detached binary system: H, L1 and L2 denote eddies associated with a
high/low pressure, respectively. Typical velocity vectors are shown with
numbers.}
\label{fig1}
\end{figure}

\subsection{Observational implication of the surface flow on the companion
star}

Davey \& Smith (\cite{dav92}) found an asymmetric irradiation pattern in
the secondary star of dwarf novae (e.g. IP Peg); the trailing hemisphere has stronger Na I line
absorption than the leading hemisphere (see also Dhillon \& Watson \cite{dhi00}). They considered that this asymmetry could be due to the
circulation current on the companion star (but see Watson et al. \cite{wat03}).

As was stated above, Oka et al. (\cite{oka02}) found three types of eddies
on the companion star. They considered that the H-eddy as well as the
L1-eddy could transport the heat due to the irradiation to the leading
hemisphere, and argued that these currents naturally explained the asymmetry
of the irradiation pattern. Note, however, that magnetic CVs, on the other
hand, show the irradiation to be on the trailing hemisphere (Davey \& Smith 
\cite{dav92},\cite{dav96}; Watson et al. \cite{wat03}). To
understand the irradiation pattern of magnetic CVs, we need another mechanism,
which could lead to the effective 
heating of the trailing hemisphere.

\subsection{Supersoft X-ray sources}

Supersoft X-ray sources (SSS) consist, like CVs, of a Roche
lobe filling companion star and a white dwarf. 
X-rays are believed to be produced by a steady nuclear burning occurring on the
surface of the white dwarf. This requires a high mass accretion
rate (${\ga} 10^{-7} M_{\odot}/$yr) that is only
possible with a high mass companion star 
(van den Heuvel et al. \cite{van92}).

In our galaxy four SSSs are known (see e.g. G\"ansicke et al. \cite{gan00}); 
one of which is RX J0019.8+2156. This
object and its 12.2 mag optical counterpart were discovered by Beuermann et
al. (\cite{beu95}). Recent observations revealed a low radial velocity
component to the He II ${\lambda}$4686 emission line (e.g. Becker et al. \cite{bec98}; Cowley et al. \cite{cow98}; Deufel et al. \cite{deu99}; McGrath et
al. \cite{mcg01}). These authors considered that the He II ${\lambda4686}$
emission line was emitted from the white dwarf or from the inner part of the
accretion disc. With this assumption they estimated the mass of the
companion to be smaller than the white dwarf. However, this estimate
certainly contradicts the massive companion theory developed by van den
Heuvel et al. (\cite{van92}).

Hachisu \& Kato (2003), on the other hand, considered that the low
radial velocity of the He II ${\lambda }$4686 emission line was emitted from the
inner hemisphere of the companion star. The inner hemisphere is irradiated
by the companion and becomes hot enough to emit the He II ${\lambda}$4686 
line. The Doppler map of He II ${\lambda 4686}$ observed by Deufel et al. (\cite{deu99}) showed that the emission line peak 
is located around $V_{\rm x}{\sim }~100$ km/s and $V_{\rm y}{\sim}~100$ km/s 
(Note that the plus sign of the above velocity
is due to our choice of the coordinate frame). 
This emission line peak does not correspond to the region
of the white dwarf or the inner part of the accretion disc. With the
irradiated companion model Hachisu \& Kato (2003) argued that such
low radial velocity emission lines were naturally explained by the massive
companion star in a system with a mass ratio of $q{\sim }3-4$ and a 
white dwarf of $0.6M_{\odot}$.

Matsumoto \& Mennickent (\cite{matm00}) constructed the Doppler
map of He II ${\lambda4686}$ of RX J0925.5-4758 which is another known Galactic SSS and also found a peak located around $V_{\rm x}\sim 100$
km/s and $V_{\rm y}\sim 100$ km/s (in our velocity sign): i.e., the low radial velocity
region. 
The low velocity of the He II ${\lambda4686}$ emission line
seems therefore to be a common feature of supersoft X-ray sources.

Hachisu \& Kato (2003) tried to interpret the low radial velocity
component of the He II ${\lambda4686}$ emission line in the Doppler map of RX
J0019.8+2156. However, their numerical model was two-dimensional, and
could therefore not handle flows on the surface of the companion.
A three-dimensional calculation is necessary in this respect, and this is the
aim of the present research.

This paper is organized as follows. In Sect. 2 we describe the assumptions of
the model and the numerical method. In Sect. 3 we show the flow patterns
and construct Doppler maps. In Sect. 4 we discuss the problem of the supersoft
X-ray sources. A discussion and a summary are given in Sect. 5.


\begin{table}
\caption{Mass ratio, computational region, and number of grid points of models}
\label{Tab:Simu}
\begin{tabular}{cccc}
\hline
Mass ratio &  & Computed region & Number of grid points \\ \hline\hline
0.05 &  & 1.0 $\times$ 1.0 $\times$ 0.5 & 101 $\times$ 101 $\times$ 51\\ 
0.33 &  & 1.0 $\times$ 1.0 $\times$ 0.5 & 101 $\times$ 101 $\times$ 51\\ 
0.5 &  & 1.1 $\times$ 1.0 $\times$ 0.5 & 111 $\times$ 101 $\times$ 51\\ 
1 &  & 1.2 $\times$ 1.0 $\times$ 0.5 & 121 $\times$ 101 $\times$ 51\\ 
2 &  & 1.4 $\times$ 1.2 $\times$ 0.6 & 141 $\times$ 121 $\times$ 61\\ 
3 &  & 1.4 $\times$ 1.2 $\times$ 0.6 & 141 $\times$ 121 $\times$ 61\\ 
\hline
\end{tabular}
\end{table}


\section{Assumptions and Numerical Method}

\subsection{Assumptions}

The binary system considered here consists of a primary compact star with
mass $M_1$ and a companion star with mass $M_2$. The system rotates
counter-clockwise viewed from the north. The mass ratio of the binary system
is defined by $q=M_2/M_1$.

We solve the three-dimensional Euler equations in the rotating frame of the
binary. The equation of state considered here is that of an ideal gas
characterized by a specific heat ratio $\gamma$. Complex effects such as
physical/turbulent viscosity (except numerical one), magnetic fields, and
irradiation from the accretion disc or the primary star are neglected.

The equations are normalized with length by the separation $A$ between the
centres of two stars, and time by $1/\Omega$, where $\Omega$ denotes the
orbital angular velocity of the binary system and, therefore, the orbital period
equals $2\pi$. The density on the inner (numerical) boundary is taken to be
unity. The gravitational constant $G$ is eliminated using the above
normalization.

\subsection{Numerical method}

We use a Cartesian coordinate system. The origin of the coordinate is
located at the centre of the primary star. The x-axis coincides with the
line joining the centres of the two stars. We define the $x-y$ plane as the orbital
plane, so that the $z$-axis is perpendicular to the orbital plane and is oriented
in the same direction as the angular momentum vector of the orbital
rotation. Note that we put the mass-donor star in the negative $x$
domain following our convention; other authors seem to prefer to put it in the
positive $x$ domain. Therefore, our sign of velocity is different from the convention
of many other authors.

The computational region is a rectangular box, and its size and the number
of grid points depend on the cases as described below. We assume symmetry about
the orbital plane, and thus only the upper half of the region is computed. We use the
simplified flux splitting (SFS) technique proposed by Jyounouchi et al. (\cite{jou93}) and Shima \& Jyounouchi (\cite{shi94}) as a Riemann solver,
and the MUSCL-type technique is used as an interpolation. The method is the
same as in Makita et al. (\cite{mak00}), Matsuda et al. (\cite{mat00}),
Fujiwara et al. (\cite{fuj01}) and Oka et al. (\cite{oka02}). The spatial
and temporal accuracy are kept at second-order levels.

\subsection{Boundary and initial conditions}

We apply the same boundary conditions as in Oka et al. (\cite{oka02}). The
inner boundary is assumed to be an equipotential surface slightly smaller
than the critical Roche lobe ($\sim 90\%$ of the critical Roche lobe radius)
of the companion. Note that this inner boundary does not necessarily mean
the real surface of the companion, but is chosen for numerical reasons.

The inside of the companion star is filled with a gas with zero velocity,
density $\rho_0=1$, and sound speed $c_0$, where $c_0$ is a free parameter.
We mainly examine the case of $c_0=0.05$. The gas can naturally flow out
from the inner boundary if the gas pressure below the inner boundary is
higher than above it.

The outside of the outer boundary is assumed to be always filled by a
gas of fixed velocity, $0$, density, $\rho_1$, and sound speed $c_1$, where $\rho_1$ and $c_1$ are parameters that are taken to be $10^{-5}$ and $\sqrt{10}$, respectively.
We note that for the calculation of the accretion discs with q=1, 2, and 3, 
the density of 
the initial ambient gas 
($\rho_1=-5$) is somewhat large, and therefore affects the accretion discs. 
In order to reduce the effect of the initial ambient gas we use a much lower ambient gas density 
of $\rho_1=-7$ when $q$ is $1$, $2$, and $3$.

At the initial time $t=0$, the entire region, except the inside of the inner
boundary, is occupied by a gas with velocity $0$, $\rho_1$, and $c_1$. We
follow the time evolution until a steady state is reached. We describe only
steady states in the following.

\begin{figure}
\centering  
\resizebox{\hsize}{!}{\includegraphics{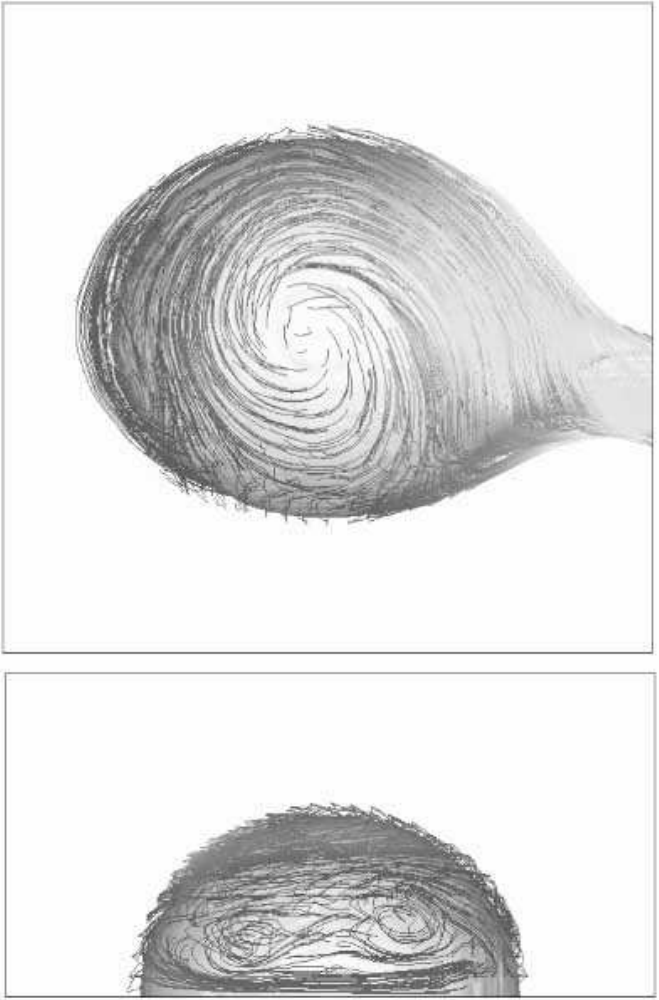}}  
\caption{Equidensity surface ($\log \protect\rho=-2$) of the companion
star and the streamlines starting from the equidensity surface of $\log 
\protect\rho=-2.5$; A mass ratio of $q=0.33$ and a specific heat ratio of $\protect\gamma=5/3$, are adopted. The top panel is viewed from the north,
while the bottom panel is viewed from the opposite side of the L1-point.
H-, L1-, and L2-eddies can be seen. }
\label{fig2}
\end{figure}

\begin{figure}
\centering 
\resizebox{\hsize}{!}{\includegraphics{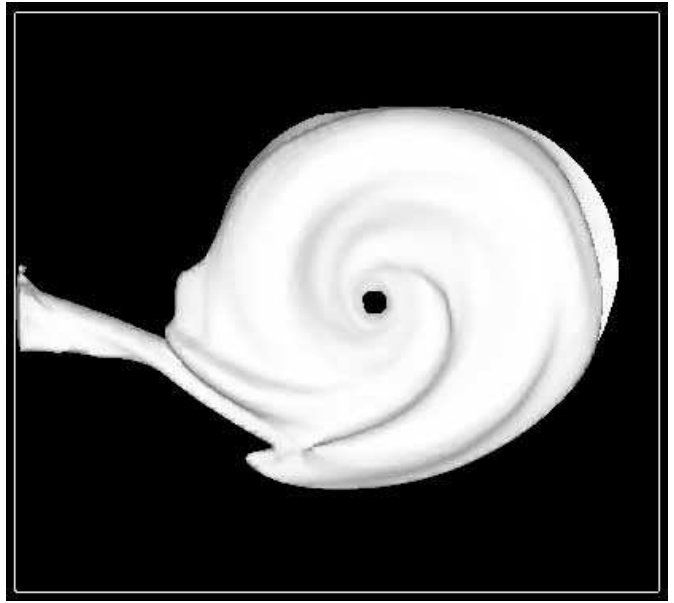}} 
\caption{Equidensity surface ($\log \protect\rho=-3.1$) of the accretion disc 
is shown; $q=0.33$ and $\protect\gamma=1.01$ are adopted in this figure. The size 
of the whole computed region is $2.0 \times 1.0 \times 0.5$ (including companion 
star), and the number of grid points is $201 \times 101 \times 51$. Note, 
however, that the region to the right of the L1 point is shown to exhibit only 
the accretion disc. We may observe a pair of spiral shocks and a shock 
associated with the L1 stream.}
\label{fig3}
\end{figure}

\section{Numerical Results}

\subsection{Surface flow}

In this section we describe the pattern of the surface flow of a companion
star. The cases of $q=0.05, 0.33, 0.5, 1, 2$, and $3$ are investigated. The
coordinate sizes and number of grid points for each case are listed in Table
1. Note that the case of $q=1$ was already studied by Oka et al. (\cite{oka02}).

For the calculation of the surface flow we adopt $\gamma=5/3$, i.e. we assume 
adiabatic expansion of the gas in the companion's atmosphere. For the calculation of 
the accretion discs, on the other hand, we use $\gamma=1.01$ (almost isothermal) 
to mimic radiation cooling, necessary to form an accretion disc.

Figure 2 shows an equidensity surface of the companion star and streamlines
originating from another equidensity surface, for $q=0.33$ and $c_0=0.05$. We can see the H-, L1-, and L2-eddies as described above;
we refer to Oka et al. (\cite{oka02}) for the mechanism of generation of these
eddies. It is remarkable that the flow pattern is essentially the same as the
case of $q=1$ in Oka et al. (\cite{oka02}). These eddies are the
manifestation of the astrostrophic wind predicted by Lubow \& Shu (\cite{lub75}).

Figure 3 depicts the equidensity surface of the accretion disc with $%
c_0=0.05$. We can see a pair of spiral shocks as well as a shock associated with 
the L1 stream.
Figures 4 and 5 represent the case of $q=2$ with $c_0=0.05$. 
Compared with Figs. 
2 and 3, we observe that the flow pattern is essentially the same as for the lower $q$ case.

Figure 6 shows the result of the extreme mass ratio, $q=0.05$ with $c_0=0.02$. 
In this case the gravitational potential due to the companion is so shallow that 
a small fraction of the gas escapes through the L2 point. We do not 
discuss here this case in detail. We can conclude that the flow pattern does not 
heavily depend on the mass ratio, $q$, at least for $q$ greater or of the order of 1.

We note an interesting possibility. As can be seen in Figs. 3 and 5, the equidensity surface rises abruptly behind spiral shocks and is thus making cliffs. These cliffs may be irradiated by the radiation emitted from the central star/the central region of the accretion disc, and thus produce spiral patterns in the Doppler map. We stress here that the spiral shock is a natural consequence of supersonic flow in a non-axisymmetric gravitational potential.

\begin{figure}
\centering  
\resizebox{\hsize}{!}{\includegraphics{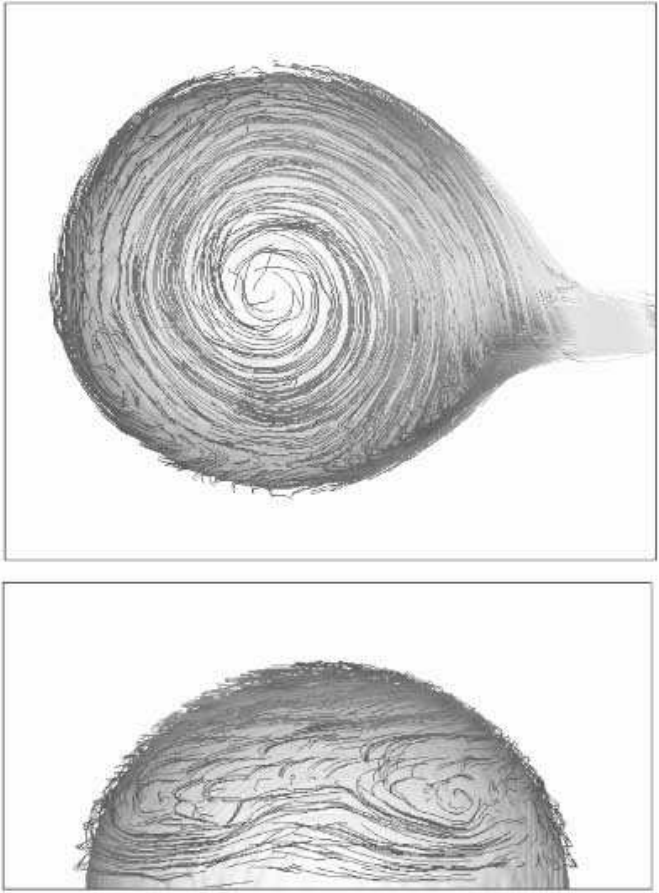}}  
\caption{Same as in Fig. 2 but for a mass ratio of $q=2$.}
\label{fig4}
\end{figure}

\begin{figure}
\centering  
\resizebox{\hsize}{!}{\includegraphics{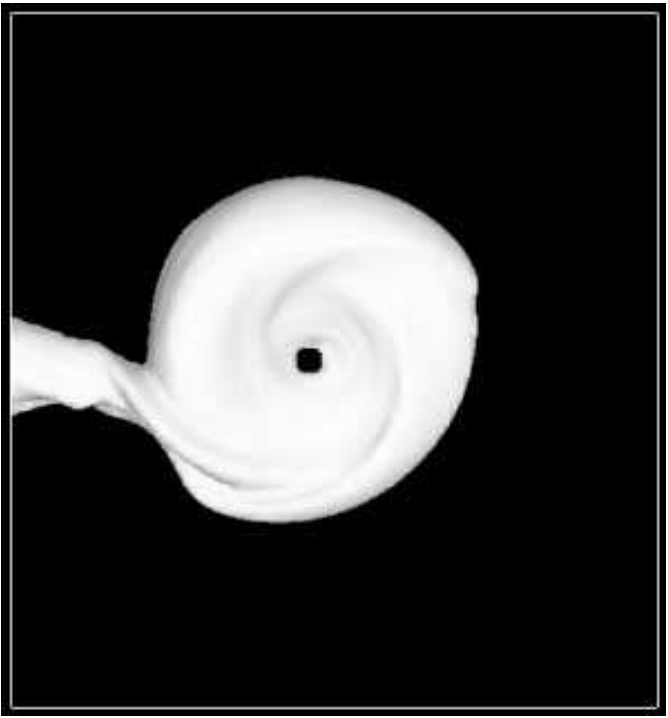}}  
\caption{Same as in Fig. 3 but here the equidensity surface of 
$\log \protect\rho=-4.5$ is shown for a mass ratio of $q=2$.}
\label{fig5}
\end{figure}

\begin{figure}
\centering  
\resizebox{\hsize}{!}{\includegraphics{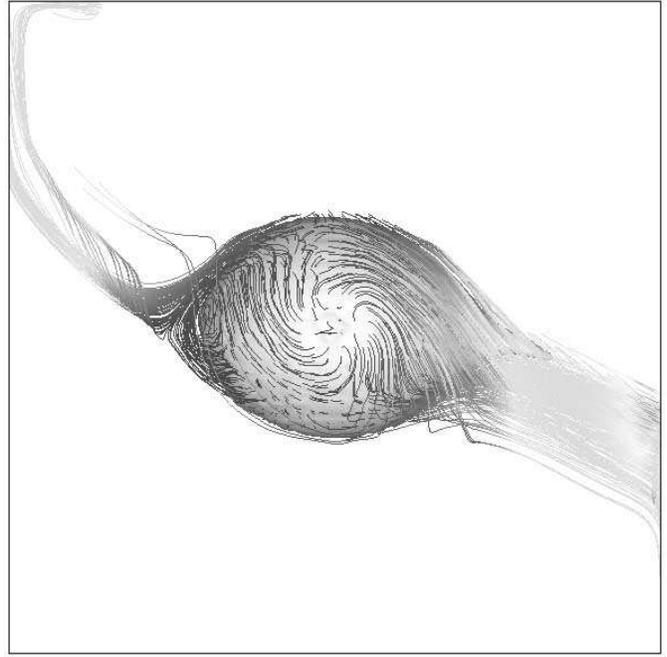}}  
\caption{Surface flow on the companion star in a binary with $q=0.05$;
equidensity surface ($\log \protect\rho=-2.5$) of the companion star and the
streamlines starting from the equidensity surface of $\log \protect\rho=-3
$. Note that a small fraction of the gas flows through the L2 point as well as 
the L1 point because of the shallow gravitational potential.}
\label{fig6}
\end{figure}

\begin{figure}
\centering  
\resizebox{\hsize}{!}{\includegraphics{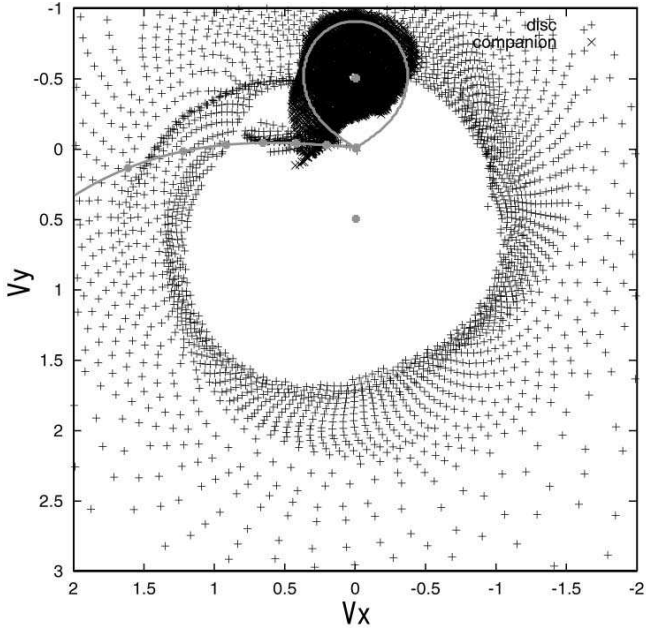}}  
\caption{Doppler map constructed from the horizontal components of velocity, 
$V_{\rm x}$ and $V_{\rm y}$, at the equidensity surface of 
$\log \protect\rho=-2.5$ for
the companion star (denoted by the cross) and 
$\log \protect\rho=-3.9$ for the accretion disc (denoted by the plus). 
In this figure $q=1$. Note that
the following figures are placed upside down in order to compare our results
with those of other authors. The critical Roche surface and the ballistic orbit of the
L1 stream are also shown for the purpose of reference. The upper and the
lower dot show the centre of mass of the companion and that of the
primary, while the middle dot (at the L1 point) denotes the common centre of
gravity.}
\label{fig7}
\end{figure}

\begin{figure}
\centering  
\resizebox{\hsize}{!}{\includegraphics{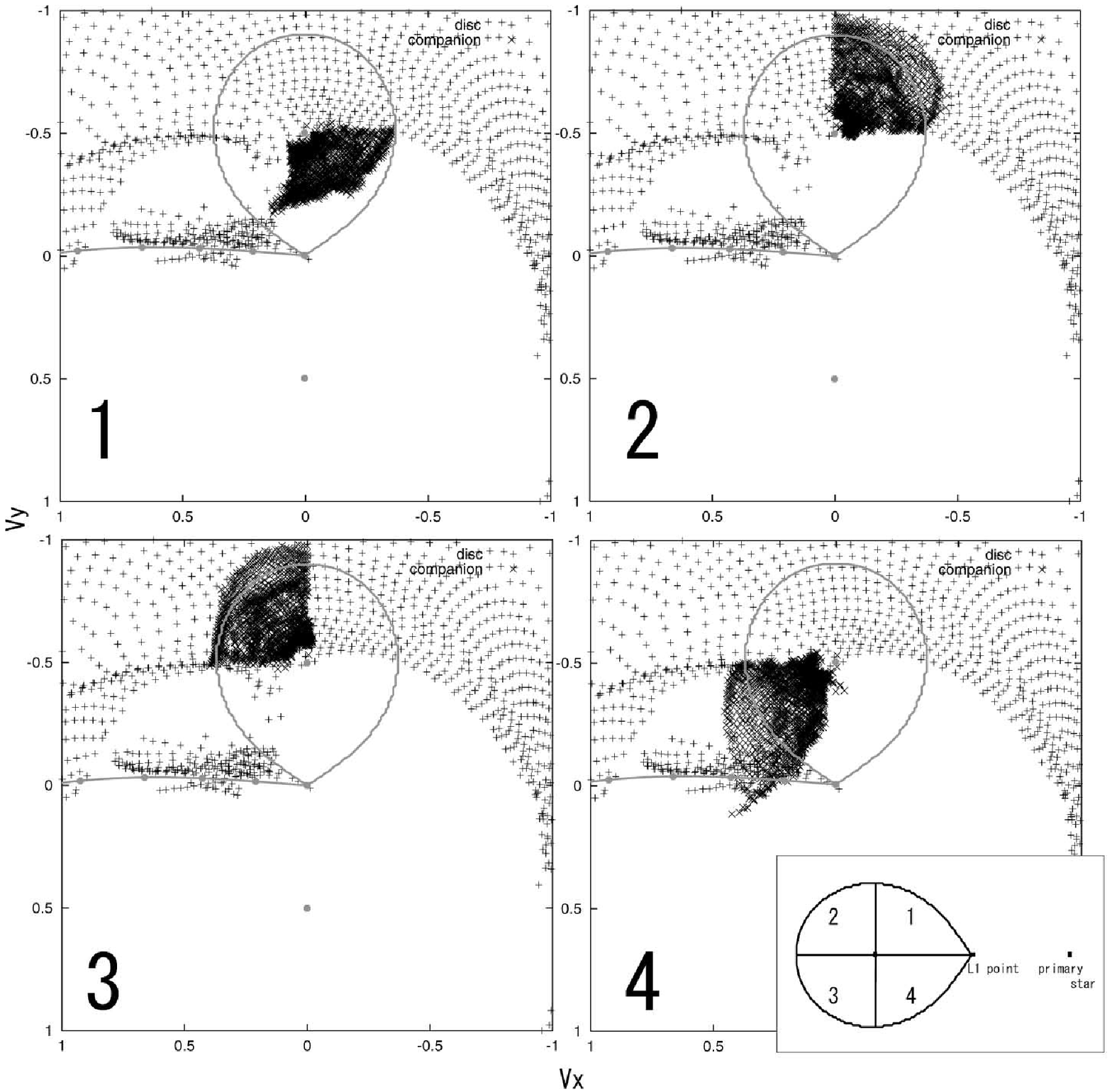}}  
\caption{The surface of the companion is divided into four quadrants as shown
in the right bottom sub-panel. We can examine the contribution from each
surface region. The number in each Doppler map corresponds to the quadrant in
the sub-panel. Note that the same data is used as in Fig. 7.}
\label{fig8}
\end{figure}

\subsection{Construction of Doppler maps}

To construct observable Doppler maps, information on the ionisation
state of atoms in the photosphere is required, and the ionisation states are mainly determined by the temperature.
The temperature in the photosphere of the
companion and the accretion disc is affected by the irradiation from the
primary and the central part of the accretion disc. Since, in the present study, 
we do not take the irradiation effect into account, we cannot
construct real observable Doppler maps but, instead,  we use the following convention.

Instead of identifying a photosphere, we use an equidensity surface as a 
possible candidate for the photosphere. The horizontal component of gas velocity, 
$V_{\rm x}$ and $V_{\rm y}$, of each computational cell at that surface is mapped onto a 
$V_{\rm x}-V_{\rm y}$ plane: the Doppler map. We tested various values of the density and found 
that the general trend was not affected by a particular choice of the 
density. 

Our construction method of the Doppler map does not reflect the 
mechanism of line emission, and therefore, only a part of our constructed 
Doppler map can be observed. For instance the outer (non-irradiated) side of the 
Roche surface may not be observable, if appropriate lines are not emitted from 
there.

Another problem in constructing a Doppler map is the choice of $\gamma$. We
use the case of $\gamma=5/3$ to calculate the contribution from the
companion, while we use the case of $\gamma=1.01$ for the accretion disc.
Thus, we combine two different simulations into one Doppler map. Therefore,
we use two different values of density on an equidensity surface: a surface
of the companion star and one for the accretion disc. Since our main aim is
to investigate Doppler maps caused by the surface flow on the companion,
the contribution from the accretion disc should here only be considered as a reference.

Figure 7 shows the constructed Doppler map of the equidensity surface of $\log \rho=-2.5$ for the companion star and $\log \rho=-3.9$ for the
accretion disc. The mass ratio is $q=1$ in this case. Note that all the Doppler 
maps shown in the figures are placed upside-down to accommodate the more 
common convention used by observers. The critical Roche
lobe of the companion star and the ballistic orbit of the L1 stream are also
shown for reference purposes. The crosses show the contribution from the
companion, while the plus signs show the contribution  from the accretion disc, which is
just for reference as was described above and we do not give a detailed
discussion about this contribution in the present paper. The contribution due to
the accretion disc collides with that of the surface flow in the Doppler map.
However, this does not necessary mean that the flow in the accretion disc
really collides with the companion in the configuration space.

The most important feature in our calculated Doppler maps is the 
asymmetry about the $V_{\rm y}$-axis. This is due to the surface flow of the
companion. If there was no flow on the companion surface, all crosses
should reside within the critical Roche surface. However, this is not the
case. We can observe that the crosses {\it deviate} to the upper right and to
the lower left.

In order to see this more clearly, Fig. 8 shows the contribution in the Doppler 
map from each quadrant of the surface region, described by the numbers in Fig. 8. 
The number in each Doppler map corresponds to the quadrant labeled by the same 
number in the right bottom sub-panel. One may safely speculate that the 
asymmetries seen in quadrant 1 and in quadrant 4 are due to the flow in 
the L1-eddy. The deviations seen in quadrants 2 and 3 are due to the L2-eddy, although 
this portion of the Doppler map is unobservable because of lack of irradiation.

We may summarize the characteristics of the Doppler map due to the companion
as follows:

\begin{enumerate}
\item The velocity vector denoted by 1 in Fig. 1 has a contribution of
positive $V_{\rm x}$, $V_{\rm y}$. This produces the asymmetries.

\item The velocity vector denoted by 2 has a contribution of positive 
$V_{\rm x}$
and negative $V_{\rm y}$.

\item The velocity vector denoted by 6 has a contribution of negative $V_{\rm x}$
and $V_{\rm y}$.

\item The velocity vector denoted by 7 has a contribution of positive 
$V_{\rm x}$ and negative $V_{\rm y}$.

\item Velocity vectors associated with the H-eddy, i.e. 3, 4 and 5, have a
tendency to shrink the cross dots in the companion towards the centre of the
companion.
\end{enumerate}

\section{Supersoft X-ray sources}
 
We apply our results to the Galactic supersoft X-ray source: RX J0019.8+2156 and try to attribute the low radial velocity component of the emission lines of He II ${\lambda}4686$ in the Doppler map observed by Deufel et al. (\cite{deu99}) to the irradiated spot on the surface of the companion.

Figure 9 shows our constructed Doppler map with $q=3$, as a model for the
RX J0019.8+2156. We assume the system velocity to be $A \Omega_{\rm orb}=$ 330km/s. The equidensity surface is chosen with $\log \rho=-2.5$ for the companion
and 
$\log \rho=-4.4$ for the accretion disc. The region where
a strong He II ${\lambda}4686$ emission line peak around $V_{\rm x}{\sim}~100$ km/s and 
$V_{\rm y}{\sim}~100$ km/s (with our velocity convention) is observed by Deufel et al. (\cite{deu99}), is denoted by a dotted circle. 

From Fig. 9, we may conclude that the case of $q\sim3$ can naturally
explain the low radial velocity component of the emission lines of He II ${\lambda}4686$. This is the same result as obtained by Hachisu \& Kato (2003). 
The mass ratio, $q\sim3$, means that the mass of the companion
star is larger than that of the white dwarf. Considering that a mass above 
$\sim0.8M_{\odot}$ for the WD is ruled out by its relatively low X-ray flux of 
$\sim0.4\times10^{37}$ergs/s (Beuermann et al. \cite{beu95}; Meyer-Hofmeister et 
al. \cite{mey98}), we may assume the mass of the white dwarf to be $\sim 
0.6M_{\odot}$. Therefore, the mass of the companion star can be estimated to be 
$\sim 2M_{\odot}$.

Finally, as was discussed earlier, the emission peak as denoted by the dotted circle 
is due to the L1-eddy, and if so, we may say that the Doppler map obtained by 
Deufel et al. (\cite{deu99}) exhibits the trace of the surface flow on the 
companion star.

\begin{figure}
\centering  
\resizebox{\hsize}{!}{\includegraphics{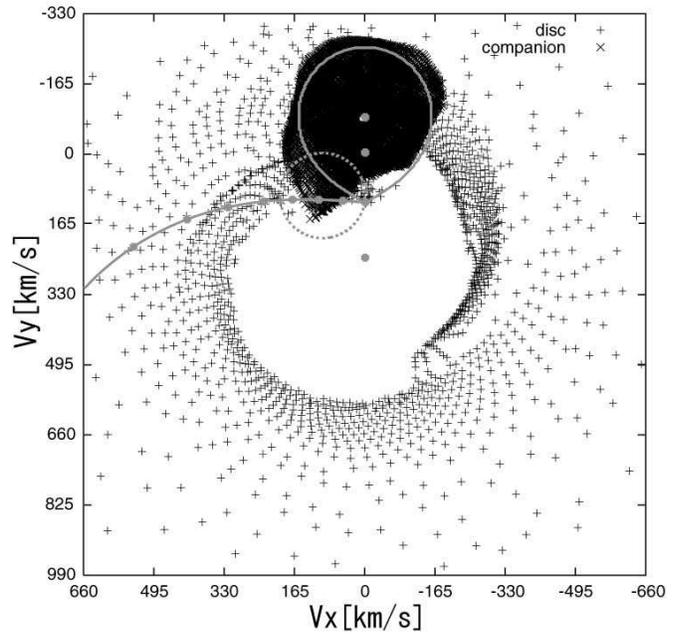}}  
\caption{Constructed Doppler map as a model of the RX J0019.8+2156 with $q=3$: in order
to have a velocity scale, we assume $A\Omega_{\rm orb}=330$ km/s. The observed
intensity of the He II ${\protect\lambda}4686$ emission by Deufel et al. 
(\cite{deu99}) has a peak around $V_{\rm x}\sim 100$ km/s and $V_{\rm y}\sim 100$ km/s (in our 
velocity sign), denoted here by a dotted circle.}
\label{fig9}
\end{figure}

\section{Summary}

We investigate the surface flow of the companion star in a semi-detached
binary system and construct model Doppler maps. Six cases with mass
ratios of $q=0.05, 0.33$, $0.5$, $1$, $2$, and $3$ are considered. We apply
our results to the Galactic supersoft X-ray source RX J0019.8+2156. The
results are summarized as follows:

\begin{enumerate}
\item We obtain the H-, L1-, and L2-eddies, which were found by Oka et
al. (\cite{oka02}), for all mass ratios. We confirm that the flow pattern does 
not heavily depend on the mass ratio.

\item We construct model Doppler maps, and we find an asymmetric velocity
distribution, due to the surface flow.

\item The companion mass of RX J0019.8+2156 is estimated to be around $2
M_{\odot}$. This massive companion is consistent with the theoretical model
proposed by van den Heuvel et al. (\cite{van92}).

\item We argue that the Doppler map of supersoft X-ray sources shows
indications of the surface flow on the companion star.
\end{enumerate}


\begin{acknowledgements}

The authors would like to thank an anonymous referee for useful 
suggestions and comments.
K.O. was supported by the Research Fellowships of the Japan Society for 
Promotion of Science for Young Scientists. T.M. was supported by the grant in 
aid for scientific research of Japan Society of Promotion of Science (13640241). 
This work was supported by "The 21st Century COE Program of Origin and Evolution 
of Planetary Systems" in Ministry of Education, Culture, Sports, Science and 
Technology (MEXT). Calculations were carried out on the SGI Origin 3800 at the 
Information Science and Technology Center of Kobe University.

\end{acknowledgements}



\begin{thebibliography}{1998}
\bibitem[1998]{bec98} Becker, C. M., Remillard, R. A., Rappaport, S. A., \&
McClintock, J. E. 1998, \apj, 506, 880

\bibitem[1995]{beu95} Beuermann, K., Reinsch, K., Barwig, H., Burwitz, V., de Martino, D., Mantel, K.-H., Pakull, M. W., Robinson, E. L., Schwope, A. D., Thomas, H.-C., Truemper, J., van Teeseling, A., Zhang, E. 1995, A\&A, 294, L1

\bibitem[1998]{cow98} Cowley, A. P., Schmidtke, P. C., Crampton, D., \&
Hutchings, J. B. 1998, \apj, 504, 854

\bibitem[1992]{dav92} Davey, S. C., \& Smith, R. C. 1992, \mnras, 257, 476

\bibitem[1996]{dav96} Davey, S. C., \& Smith, R. C. 1996, \mnras, 280,481

\bibitem[1999]{deu99} Deufel, B., Barwig, H., \v{S}imi\'{c}, D., Wolf, S., \&
Drory, N. 1999, A\&A, 343, 455

\bibitem[2000]{dhi00} Dhillon, V. S., \& Watson, C. S. 2000, Proc. of the
Astro-tomography Workshop, Brussels, July 2000, ed. H. Boffin, \& D. Steeghs
(Springer-Verlag Lecture Notes in Physics), 94

\bibitem[2001]{fuj01} Fujiwara, H., Makita, M., Nagae, T., \& Matsuda, T.
2001, Progr. of Theor. Phys., 106, 729
 
\bibitem[2000]{gan00} G\"ansicke , B. T., van Teeseling, A., Beuermann, K., \& Reinsch, K. 2000, New Astr. Rev., 44, 143



\bibitem[2003]{Hach03} Hachisu, I., \& Kato, M. 2003, ApJL, submitted

\bibitem[1993]{jou93} Jyounouchi, T., Kitagawa, S., Sakashita, S., \&
Yasuhara, M. 1993, Proc. 7th CFD Symp.

\bibitem[1975]{lub75} Lubow, S. H., \& Shu, F. H. 1975, \apj, 198, 383

\bibitem[2000]{mak00} Makita, M., Miyawaki, K., \& Matsuda, T. 2000, \mnras,
319, 906

\bibitem[2000]{mat00} Matsuda, T., Makita, M., Fujiwara, H., Nagae, T.,
Haraguchi, K., Hayashi, E., \& Boffin, H. M. J. 2000, \apss, 274, 259

\bibitem[2000]{matm00} Matsumoto, K., \& Mennickent, R. E. 2000, A\&A, 356,
579

\bibitem[2001]{mcg01} McGrath, T. K., Schmidtke, P. C., Cowley, A. P.,
Ponder, A. L., \& Wagner, R. M. 2001, \aj, 122, 1578

\bibitem[1998]{mey98} Meyer-Hofmeister, E., Schandl, S., Deufel, B., Barwig,
H., \& Meyer, F. 1998, A\&A, 331, 612

\bibitem[2002]{oka02} Oka, K., Nagae, T., Matsuda, T., Fujiwara, H., \&
Boffin, H. M. J. 2002, A\&A, 394, 115


\bibitem[1994]{shi94} Shima, E., \& Jyounouchi, T. 1994, NAL-SP27, Proc. of
12th NAL Symp. on Aircraft Computational Aerodynamics, 255


\bibitem[1992]{van92} van den Heuvel, E. P. J., Bhattacharya, D., Nomoto,
K., \& Rappaport, S. 1992, A\&A, 262, 97

\bibitem[2003]{wat03} Watson, C. A., Dhillon, V. S., Rutten, R. G. M., \&
Schwope, A. D. 2003, \mnras, 341, 129
\end{thebibliography}
\end{document}